\documentclass[12pt]{article}
\usepackage{epsf}
\def\beq{\begin{equation}}
\def\eeq{\end{equation}}
\def\ap#1#2#3 {Ann. Phys. (NY) {\bf#1} (19#2) #3}
\def\err#1#2#3 {{\it Erratum} {\bf#1} (19#2) #3}
\def\ib#1#2#3 {{\it ibid.} {\bf#1} (19#2) #3}
\def\ijmp#1#2#3 {Int. J. Mod. Phys. {\bf#1} (19#2) #3}
\def\jetp#1#2#3 {JETP Lett. {\bf#1} (19#2) #3}
\def\mpl#1#2#3 {Mod. Phys. Lett. {\bf#1} (19#2) #3}
\def\np#1#2#3 {Nucl. Phys. {\bf#1} (19#2) #3}
\def\pl#1#2#3 {Phys. Lett. {\bf#1} (19#2) #3}
\def\prep#1#2#3 {Phys. Rep. {\bf#1} (19#2) #3}
\def\prev#1#2#3 {Phys. Rev. {\bf#1} (19#2) #3}
\def\prl#1#2#3 {Phys. Rev. Lett. {\bf#1} (19#2) #3}
\def\sjnp#1#2#3 {Sov. J. Nucl. Phys. {\bf#1} (19#2) #3}
\def\spj#1#2#3 {Sov. Phys. JETP {\bf#1} (19#2) #3}
\def\spu#1#2#3 {Sov. Phys. Usp. {\bf#1} (19#2) #3}
\def\zp#1#2#3 {Zeit. Phys. {\bf#1} (19#2) #3}

\begin{document}
\begin{titlepage}
\begin{center}
{\Large \bf Theoretical Physics Institute \\
University of Minnesota \\}  \end{center}
\vspace{0.2in}
\begin{flushright}
TPI-MINN-01/48-T \\
UMN-TH-2030-01 \\
November 2001 \\
\end{flushright}
\vspace{0.3in}
\begin{center}
{\Large \bf  More remarks on suppression of large black hole production in particle collisions
\\}
\vspace{0.2in}
{\bf M.B. Voloshin  \\ }
Theoretical Physics Institute, University of Minnesota, Minneapolis,
MN
55455 \\ and \\
Institute of Theoretical and Experimental Physics, Moscow, 117259
\\[0.2in]
\end{center}

\begin{abstract}

The problem of evaluating the cross section of production of large black holes in particle collisions at trans-Planckian energies is revisited. It is argued that the geometric cross section claimed in the literature would in fact lead to an exponentially growing with energy total cross section dominated by production of many ``small" black holes. On the other hand, the semiclassical treatment of this problem, leading to an exponentially suppressed cross section for production of large black holes, requires that in the classical limit the cross section vanishes for massless colliding particles. The latter behavior can in principle be probed by numerical simulations in the classical general relativity.

\end{abstract}
\end{titlepage}

\section{Introduction}

In the standard theory of gravity the problem of particle collisions at trans-Planckian energies, i.e. far above the Planck energy scale, has no chance to be tested experimentally or by observation of cosmic rays. However this problem has recently attracted a lively attention in connection with models with extra spatial dimensions, where the multi-dimensional analog of the Planck energy scale can be as low as 
few TeV \cite{add}, and thus may be within the reach of realistic accelerators and also within the range of energies of cosmic rays. 

It has been claimed\cite{bf,gt,dl} (see also \cite{de,sg}) that the dominant process in such collisions should be a collapse of a finite part of the total c.m. energy of the colliding particles $\sqrt{s}$ into a black hole, whose mass should thus be $M_H \sim \sqrt{s}$. The cross section for this process is claimed to be given by the geometric cross section area of the black hole horizon with the radius $r_H$, $\sigma(2 \to BH) \sim \pi \, r_H^2$. Some consequences of this claimed behavior of the cross section for black hole production  within models with extra dimensions were recently considered  for cosmic ray observations\cite{fs,ag,rt} and for experiments at future colliders\cite{kc}.

The reasoning for the geometric classical cross section is based on two points. The first being that for collisions at the impact parameter $b$ equal to zero (axisymmetric head-on collisions) in a (2+1) dimensional anti-de-Sitter space it has been shown\cite{hjm} that a black hole is inevitably created, while in (3+1) dimensions the estimates of the radiated away gravitational energy\cite{pdd} leave a finite fraction of the total energy of the colliding particles reaching small separations, and plausibly forming a large black hole with mass $M_H \sim \sqrt{s}$. The second point in the reasoning is that a natural scale for the impact parameter $b$ in off-axis collisions is given by the gravitational radius $r_H$. Thus an extrapolation of the behavior at $b=0$ to such range of the impact parameter yields the claimed estimate $\sigma(2 \to BH) \sim \pi \, r_H^2$\,\footnote{This reasoning is most clearly formulated by Banks and Fischler\cite{bf}: ``General Relativity predicts that when the impact parameter  \ldots \ is smaller than a critical value $R_S$, a black hole is formed. \ldots $R_S$ is of order the Schwarzchild radius of the corresponding black hole\ldots". Notably, no supporting references or arguments are supplied with these statements.}.

These estimates were argued in Ref.\cite{mv} to be grossly incorrect, and two semiclassical approaches to the problem were presented, one based on a path integral calculation of the production amplitude, and the other based on thermodynamic properties of black holes, both leading to the conclusion that the cross section is exponentially suppressed by the factor $\exp(-I_E)$, with $I_E$ being the Euclidean Gibbons-Hawking\cite{gh} action for the black hole. 

In view of the recent interest to the discussed problem and also in view of an explicit contradiction between the conclusion of \cite{mv} and seemingly more intuitive geometric estimates\cite{bf,gt,dl}, it is worthwhile to discuss additional ``pro and con" arguments. This paper contains two remarks relevant to the existing controversy. The first is based on considering the energy fragmentation of the colliding particles due to emission of gravitons. It is argued in Sec.2 that if one assumes the geometric cross section, growing as a power of the total energy, a fragmentation of the initial energy with subsequent formation of multiple smaller black holes should be even more probable, so that the sum of the cross section over the channels with multiple black holes should grow exponentially with energy. If one further assumes that unitarity is restored in some form, so that the total cross section is non-exponential, the partial cross section into a single large black hole has to be exponentially suppressed. The second remark to be presented in Sec.3 addresses the relation between a semiclassical calculation\cite{mv} of the exponentially suppressed cross section and a possible classical behavior. Formally, the Gibbons-Hawking exponent requires that the cross section vanishes in the limit $\hbar \to 0$. In terms of the effective impact parameter for creation of a black hole in collision of two ultrarelativistic (formally massless) particles the factor $\exp(-I_E)$ corresponds to an effective range of $b$ being independent of the energy and given by the Planck length $b \sim \sqrt{G \, \hbar}$ (in (3+1) dimensions), which also vanishes in the limit $\hbar \to 0$, leading to distinct implications for off-axis classical collisions that can possibly be verified by numerical simulations. Finally, in  Sec. 4 the conclusions of the present paper are summarized and also specific most recent comments on the argumentation of Ref.\cite{mv} are addressed.

The standard General Relativity in (3+1) dimensions is implied throughout the discussion in this paper with no further reference to schemes with extra dimensions.  

\section{Energy fragmentation and multi-black-hole production}

The geometric formula  $\sigma \sim \pi \, r_H^2$ implies that the cross section for the black hole production grows quadratically in the mass $M_H$, since $r_H=2 \, G \, M_H$ with $G$ being the Newton's constant. Clearly, at a far trans-Planckian energy $E=\sqrt{s}$, such that $G \, E^2 \gg 1$, one might find a larger probability if the energy is split in several fragments, and those fragments collide to produce several black holes of smaller masses. We discuss here such process in the case where the fragments are gravitons, and the number $n$ of produced black holes is large, $n \gg 1$, so that the typical energy $\omega \sim E/n$ of each graviton is much smaller than $E$. On the other hand, it is assumed here that the typical invariant mass in pairwise collisions of the gravitons is still larger than the Planck mass, so that one could apply the geometric formula for creation of ``small" black holes in those collisions. The latter condition allows to only consider the range of $n$ up to $n \sim \sqrt{G} \, E$. 

For the estimate of the effect of the energy fragmentation into gravitons we start with considering a single soft graviton bremsstrahlung in a collision involving ultrarelativistic particles.
%\footnote{I believe that the behavior described in this and the next %paragraphs must be well known to experts in gravitational radiation. %However I cannot point to a specific reference}. 
The  term in the amplitude, corresponding to emission of a soft graviton with momentum $\vec{k}$ by massless particle ``$a$" with energy $E_a \gg \omega$, can be simply found in the physical gauge in the c.m. frame, where the components of the graviton tensor amplitude $h_{\mu \nu}$ are only spatial, traceless, and transversal to the graviton momentum $\vec{k}$:
\beq
A_a = \sqrt{16 \pi \, G} \, {E_a \over  \omega}\, {{\hat p}^i {\hat p}^j h_{ij}  \over 1-\cos \theta}~.
\label{aa}
\eeq 
Here $i$ and $j$ stand for the spatial indices, and ${\hat p}^i$ is a unit vector in the direction of the momentum of the particle $a$, and $\theta$ is the angle between that direction and the graviton momentum. Also it is assumed that the graviton tensor amplitude is canonically normalized, i.e. $g_{\mu \nu}= \eta_{\mu \nu}+ \sqrt{16 \pi \, G} \, h_{\mu \nu}$. One can further notice that for the physical components of the graviton tensor amplitude, the product ${\hat p}^i {\hat p}^j h_{ij}$ is in fact proportional to $\sin^2 \theta$. Thus unlike in a bremsstrahlung of massless vector bosons (e.g. photons) there is no forward peak in the emission of gravitons for an ultrarelativistic particle, i.e. in the massless limit.

The total amplitude of a soft graviton emission is given by the sum of the amplitudes of emission, as in eq.(\ref{aa}), by all the energetic particles. In particular this generally leads to that, unlike in the familiar case of bremsstrahlung of vector particles, the direction of emission of soft graviton is not associated with the direction of any particular incoming or outgoing particle. This is most explicitly illustrated by the graviton bremsstrahlung in a collision of two massless particles at energy $E=\sqrt{s}$ forming a static (in the c.m.) massive object (as in the discussed production of a single large black hole). The total amplitude of emission can be written in the c.m. frame as
\beq
A=\sqrt{16 \pi \, G} \, {E \over 2 \omega}\, {\hat p}^i {\hat p}^k h_{ik} \left ( {1 \over 1-\cos \theta} + {1 \over 1+ \cos \theta} \right )~,
\label{acm}
\eeq
which results in totally isotropic probability of emission of the graviton:
\beq
dw={2 \, G \, s \over \pi} \, {d \omega \over \omega} {d \Omega \over 4 \pi}~,
\label{pcm}
\eeq
where $d \Omega$ is the differential of the solid angle. It is important for what follows that the effective strength of the source of soft gravitons is determined by the large energy $E$ of the projectile, rather than by the soft graviton energy $\omega$.

Proceeding to considering the fragmentation of the energy of the initial particles, we first discuss a condition that would ensure that the produced fragments do not subsequently fall into a common large black hole. Most conservatively, i.e. in a manner most favorable to the idea of geometric cross section for black hole production, it is assumed here that all objects moving at transverse distances shorter than the gravitational radius of the largest possible common black hole, $r_0 = 2 \, G \, \sqrt{s}$, are likely to fall into the large black hole. Thus only the fragments, that move at transverse distances larger than $r_0$ will be considered as avoiding that fall. One can readily verify that this condition limits the transverse momenta of the ``fall safe" fragments as $k_\perp < 1/r_0$. The longitudinal distance at which a fragment with energy $\omega$ is emitted can be estimated as $\ell \sim \omega/k_\perp^2$, and this distance is also larger than $r_0$, once the above condition for $k_\perp$ is satisfied. 

\begin{figure}[ht]
  \begin{center}
    \leavevmode
    \epsfbox{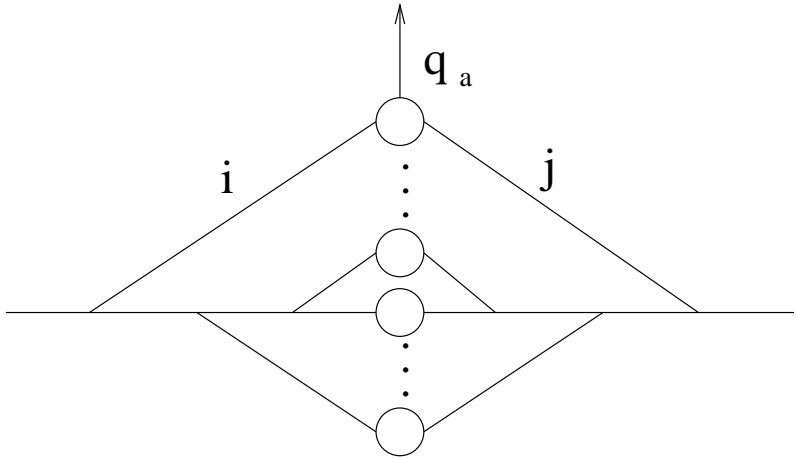}
    \caption{A representative of the type of graphs considered here for multiple black hole production. The circles stand for black holes and the lines denote gravitons. (The initial incoming particles are also drawn as gravitons for simplicity.)}
  \end{center}
\label{fig:xy}
\end{figure}

Let us estimate now the amplitude for production of $n$ black holes due to collisions of soft virtual (in fact almost real) gravitons, under the assumption of the geometric cross section. A generic graph for this process is shown in Fig.1. According to the previous discussion of soft graviton emission the factor in the amplitude describing production of black hole in the collision of $i$-th and $j$-th gravitons can be estimated as
\beq
\int \, {G \, s \over \omega_i \, \omega_j} \, f(q^2) {d^4 k_i \over k_i^2 \, k_j^2}~,
\label{bhf}
\eeq
where $f(q^2)$ is the coupling of two gravitons to a state of black hole with mass $M_H^2=q^2$. In evaluating the amplitude we treat the logarithmic integrals as being of order one, which is sufficient for estimating the lower bound on the amplitude. In this approximation the result of the integration over $k_i$ (with the restriction $k_\perp < 1/r_0$ for both $k_i$ and $k_j$) can be estimated as $\int d^4 k_i / (k_i^2 \, k_j^2 ) \sim O(1)$, and $\omega_i \, \omega_j \sim q^2 = M_H^2$. Then the cross section for producing $n$ smaller black holes can be estimated as
\beq
d \sigma_n \sim {1 \over (n!)^2} \prod_{a=1}^n {G^2\, s^2 \over (q_a^2)^2}\, |f(q_a^2)|^2 \, {d^3 q_a \over q_a^0} \, \rho(q_a^2) \, dq_a^2~,
\label{sbh}
\eeq
where the index $a$ enumerates the produced black holes, and $\rho(M_H^2)$ is the density of states of a black hole at mass $M_H$. The factor $(n!)^{-2}$ in eq.(\ref{sbh}) arises from the number ($2n$) of identical (virtual) gravitons, and we neglect weaker in $n$ factors, i.e. behaving as powers of $n$ or as $c^n$ with $c$ being a numerical constant. With the constraint $q_\perp < 1/r_0$ the integration over the momentum $q_a$ of the black hole can be estimated (again, up to a logarithmic factor) as $\int d^3q/q^0 \sim 1/r_0^2 \approx 1/(G^2 s)$. Furthermore, the geometric cross section, that is assumed here for the purpose of this calculation, implies that
\beq
|f(q^2)|^2  \, \rho(q^2) \sim G^2 \, (q^2)^2~,
\label{sdbh}
\eeq
which according to eq.(\ref{sbh}) results in the estimate of the cross section as
\beq
d \sigma_n \sim {1 \over (n!)^2} \prod_{a=1}^n {G^2\, s} \, d q_a^2~.
\label{sbhf}
\eeq

For production of $n$ black holes, each with mass ranging up to $E/n$, (the lower bound on) the cross section can thus be estimated as 
\beq
\sigma_n \sim \left ( {G^2 \, s^2 \over n^4} \right )^n~.
\label{sbht}
\eeq

In obtaining this estimate for the cross section the graphs with graviton self-interactions were neglected. The contribution of emission of gravitons by gravitons would enhance the amplitude, and eq.(\ref{sbht}) can still be used as a lower bound. A more serious problem arises from loop graphs with rescattering of gravitons. These graphs generally would modify the amplitude by order one. However a reliable estimate of the effect runs into the general problem of calculating loop graphs in quantum gravity, which is not readily solvable at present. This undoubtedly makes the status of the estimate (\ref{sbht}) less certain, although it does not look any less certain than that of the geometric cross section.

Clearly, the estimate (\ref{sbht}) implies that at $G \, s \gg 1$ the total cross section should grow exponentially $\sigma_{tot} \sim \exp(\sqrt{G \, s})$, and should be dominated by production of $O(\sqrt{G \, s})$ small black holes, each having mass of order the Planck mass $G^{-1/2}$. I believe that this behavior illustrates the intrinsic inconsistency of the assumption of a geometric cross section for black hole production. Namely, assuming the geometric formula for production of a single black hole, one arrives at the conclusion that the channel with a single black hole should make only an exponentially small fraction of the total cross section. Thus in any unitary picture, where the total cross section does not grow exponentially with energy, the partial cross section with production of one large black hole should be exponentially small, in contradiction with the assumption of geometric cross section.

\section{Semiclassical cross section, and its classical limit}

A semiclassical treatment of the black hole production in particle collisions results\cite{mv} in a formula, which is completely different from the geometric one. Namely the cross section for production of a large black hole is exponentially suppressed by the factor $\exp(-I_E)$ with $I_E$ being the (Euclidean) Gibbons-Hawking action\cite{gh} for the black hole. For a non-charged black hole with mass $M_H$ and angular momentum $J$, the production cross section is thus estimated as
\beq
\sigma(M_H,J) \sim \exp \left [ -{2 \, \pi \, G \, M_H^2 \over \hbar} \left ( 1 + {1 \over \sqrt{1-J^2/(G^2 \, M_H^4)}} \right ) \right ]~,
\label{sgh}
\eeq
where the dependence on the Planck's constant $\hbar$ is restored.
The explicit $J$ dependence in eq.(\ref{sgh}) allows one to find values of the characteristic angular momentum with which a black hole is produced:
$J \sim \sqrt{G \, \hbar} \, M_H$ and thus to estimate the range of the impact parameter $b$ at which the black hole is produced:
\beq
b = {J \over M_h} \sim \sqrt{G \, \hbar}~.
\label{ip}
\eeq
In other words, the characteristic impact parameter is of order the Planck length and does not depend on the energy of colliding particles.

Clearly, for a large black hole, $G M_H^2 \gg 1$, this estimate is much smaller than in the geometric picture: $b \ll r_H$. Moreover, in the classical limit, $\hbar \to 0$, both the impact parameter and the exponential expression in eq.(\ref{sgh}) for the cross section of the black hole production both go to zero. This undoubtedly implies that classically two ultrarelativistic particles either do not form a black hole at any $b$, even though at $b=0$ they are shown\cite{pdd} to approach a zero separation retaining a finite fraction of the initial energy, or the behavior in $b$ is singular, and  they collapse into a black hole only in a head-on collision and that such collapse does not take place in off-axis collisions with any finite offset.

The latter singular in $b$  behavior might look 
unnatural. This singularity however may well be a result of a somewhat 
singular limit usually considered in the classical setup of the 
problem\cite{pdd}. Namely, in the classical setting one considers a 
configuration, which initially is described by two Schwarzchild metrics 
boosted towards each other. The mass $m$ corresponding to each metric is 
small. Furthermore, the limit is considered in which the mass $m$ is 
taken to zero with the energy per particle $E/2=m \, \gamma$ being 
fixed. In this massless limit the boosted Schwarzchild metric is 
described by the well-known Aichelburg-Sexl shock wave\cite{as}. The 
latter metric is however quite singular, e.g. the curvature invariants 
are identically zero everywhere, except for the point of classical 
location of the particle. Thus a singular dependence on the 
impact parameter of a classical collapse into a black hole may in fact 
be merely a reflection of the singularity of the Aichelburg-Sexl metric, 
that is smoothly regularized for finite values of $m$ and $\gamma$.

\section{Discussion and some comments on recent literature}
The common physical intuition about universality of gravitational collapse of any form of energy into a black hole, which is certainly correct for non-relativistic objects, might turn to be misleading for collisions of ultrarelativistic particles. The geometric formula for the cross section of production of a large black hole, naturally following from this intuitive picture of universal collapse, gives rise to a scattering behavior that does not comply well with either physical intuition or unitarity. Namely, as argued in Sec. 2, assuming the geometric formula, one arrives at the conclusion that fragmentation of the energy of initial particles and producing multiple smaller black holes is more probable than creation of just one large black hole. The total cross section for multiple-black hole production is then exponentially larger than for one black hole, and is dominated by final states with masses of produced black hole of order Planck mass. Thus one should either accept a non-intuitive and non-unitary exponential growth of the total cross section with energy, or admit that the geometric formula is incorrect. 

The semiclassical approach\cite{mv} resulting in eq.(\ref{sgh}) with an exponentially suppressed cross section for production of large black holes, undoubtedly is at variance with the picture of universality of gravitational collapse. As discussed in Sec. 3, the $\hbar \to 0$ limit of the semiclassical formula in fact requires the classical cross section for producing a large black hole to be zero in the ultrarelativistic limit. At this point it can only be suggested as a speculation that a numerical solution of the general relativity equations can be analyzed for {\it off-axis} collisions of either two Aichelburg-Sexl solutions, or for two smoothed configurations at finite $m$ and $\gamma$, in order to either confirm or disprove the classical behavior following from eq.(\ref{sgh}). 

A vanishing classical cross section for black hole production would certainly have to be reconciled with the picture of universal collapse into a black hole. At present one may speculate on two logical possibilities for such reconciliation. One possible resolution would be that essentially all the energy of initial particles gets radiated away before they approach each other at the distance of order $r_0$, so that no collapse of energy of order $\sqrt{s}$ can take place. This behavior was also suggested in Ref.\cite{mv}, based on the observation that the exponential factor in eq.(\ref{sgh}) is somewhat analogous to Sudakov form factor in QED with the effective strength $G \, E^2$ in place of the QED coupling $\alpha$, albeit without the usual QED logarithms. However this reasoning becomes less certain upon a closer inspection. Indeed, one can estimate the total energy radiated away using the formula in eq.(\ref{pcm}). In order to include only the gravitons radiated at distances larger than $r_0$ the upper cutoff in the soft graviton energy should be set at $\omega \sim 1/r_0$, which gives the estimate of the total radiated energy as
\beq
E_{rad} \sim {G \, s \over r_0} \sim \sqrt{s}~,
\label{erad}
\eeq
and the total number of emitted gravitons is of order $G \, s$. According to this estimate the radiated energy is of order the total energy in the process\footnote{It can be noted that the back reaction of the graviton radiation gives rise to an uncertainty in the impact parameter of order $\sqrt{G \, s}/\sqrt{s} = \sqrt{G}$, i.e. of order of the Planck length, which agrees with the estimate of the effective impact parameter in eq.(\ref{ip}).}. However the accuracy of the simple one-graviton formula (\ref{pcm}) is generally insufficient to assert that exactly {\it all} the energy is radiated away. On the other hand, the total energy loss in bremsstrahlung is a classical quantity, and in principle can be found classically. According to the estimates\cite{pdd} of classical radiation not more than about 16\% of the initial energy is lost in radiation in an on-axis collision of Aichelburg-Sexl shock waves. Although, a non-trivial behavior of the total amount of radiated energy on the impact parameter in off-axis collisions is quite possible, the problem of graviton radiation certainly remains open and is subject to further study.

Another possible resolution of the contradiction with the intuitive picture of collapse can be that in an essentially non-spherical configuration the rapid time evolution of the energy density distribution prevents from the collapse into a black hole. One argument, recently suggested\cite{sg} in favor of the geometric cross section appeals to the ``Thorne's hoop suggestion\cite{thorne}, which states that horizons form when and only when a mass $M$ is compacted into a region whose circumference in every direction is less than $2\pi r_h(M)$". However,  as useful as Thorne's hoop suggestion may be for sufficiently slowly evolving mass distributions, for the case of ultrarelativistic evolution, as one encounters in trans-Planckian particle collisions, this suggestion needs at least a better formulation, as regards the issues of e.g. simultaneity in defining the ``circumference in every direction", before its status can be ascertained in this setting.

Summarizing the discussion in two previous paragraphs, it should be admitted that reconciling the possible zero cross section in the classical limit, as suggested by eq.(\ref{sgh}), with the intuitive picture of collapse undoubtedly presents a problem, which however can hopefully be resolved by classical, possibly numerical, methods.

I also use the opportunity presented by this `discussion' section to answer some specific recent comments on the argumentation presented in Ref.\cite{mv}. Recently Giddings\cite{sg} has argued that a semiclassical calculation of the black hole production should be invalid, since a Euclidean action arises only for processes that are classically forbidden. He also states: ``\ldots black hole formation is a {\it classically allowed} (in fact compulsory!) process". I believe that the discussion presented in Sec.3 and earlier in this section clearly suggests that the question whether the formation of a black hole in an {\it ultrarelativistic} collision is ``classically allowed" or forbidden is still not resolved. One of the ways of probing this issue in the absence of a good classical description is to attempt a semiclassical calculation as was done in Ref.\cite{mv}. In such calculation the Euclidean action becomes essential and suggests that classically the cross section should vanish. This argument at least indicates that a further study is needed, classical or semiclassical. 

Another criticism, expressed in Ref.\cite{sg} relates to the CPT treatment of the reciprocity between the production and decay amplitudes for the states of a black hole. Namely, it is argued that the time reversal of a black hole is a white hole "which is a very different state". It should be noted at this point that, using the picture of a black hole as being an ensemble of a very large number of resonant states\cite{gt}, the reciprocity in Ref.\cite{mv} is treated exactly as for ordinary resonances: it relates the amplitudes of formation and decay of a resonance, rather than relating a formation amplitude for a resonance to a decay amplitude of a ``very different" (and unphysical) state  whose wave function formally exponentially grows with time. A somewhat similar reasoning based on reciprocity for processes involving states of a black hole can be also found in Ref.\cite{thooft}. The thermodynamic argument might indeed have caveats of its own, as discussed  in Ref.\cite{mv}, however not related to reciprocity.

Finally, Dimopoluos and Emparan\cite{de} argued against the exponential suppression of the cross section on the basis that the geometric formula matches their estimate of the cross section of production of string-theory objects, `string-balls', at intermediate energies, where the general relativity limit is not applicable, and the string dynamics should be considered instead. However, if general relativity is replaced by string theory, or some other scheme, the Gibbons-Hawking action, used in eq.(\ref{sgh}), would also become invalid at intermediate energies around the Planck mass, and has to be replaced by appropriate expression in that more universal theory. Thus no problems with matching the cross section across the energy spectrum should arise. In any case, it is highly unlikely that at far trans-Planckian energies (e.g. of order solar mass), where the general relativity is perfectly applicable, such gross features of the discussed cross section as the difference between the geometric and semiclassical formulas would depend on details of the string theory, or some other scheme of quantum gravity. Rather the problem of large black hole production should be solvable within classical or semiclassical standard general relativity.

In conclusion. At present there is no compelling reason to believe that formation of a large black hole in particle collisions at trans-Planckian energies is a ``compulsory" process and is described by a geometric cross section, $\sigma \sim \pi \, r_H^2$. Moreover, a semiclassical treatment\cite{mv} suggests that this cross section is exponentially suppressed, and in fact vanishes in the classical ultrarelativistic limit. To an extent, this issue can be resolved by a further analysis, e.g. numerical, of ultrarelativistic collisions in the classical theory of gravity. It may well be that a further study of the fascinating problem of creating black holes, as useless as it most probably is for practical collisions, will bring new conceptual developments in understanding gravity.

\section{Acknowledgements}
I acknowledge the hospitality of the Aspen Center for Physics, where initial estimates for this paper were done.
This work is supported in part by DOE under the grant
number DE-FG02-94ER40823.

\end{document}